\documentclass[onecolumn,noshowpacs,nofootinbib,11pt]{revtex4}
\usepackage{float,xcolor,upgreek,yfonts,tikz}
\usepackage{color} 
\usepackage{textcomp,bm}
\usepackage{graphicx}
\usepackage{subfigure}
\usepackage{braket}
\usepackage{physics}

\def\b0{{\bf 0}}

\def\bdm{\begin{displaymath}}
\def\edm{\end{displaymath}}

\def\Tr{{\mathrm{Tr}}}

\newcommand{\be}{\begin{eqnarray}}
\newcommand{\ee}{\end{eqnarray}}

\renewcommand{\b}{\hat b}

\newcommand{\Det}{\text{Det}\,}
\newcommand{\del}{\partial}

\newcommand{\action}{\mathsf{S}}

\newcommand{\source}{\varphi^{(0)}}
\newcommand{\sourceG}{h^{(0)}_{xy}}

\newcommand{\bmq}{{\vec{q}}}

\newcommand{\bmx}{{\vec{x}}}

\newcommand{\ikx}{ {i \omega t - i \bmq \cdot \bmx} }

\usepackage{graphicx,float}\usepackage[all]{xy}
\usepackage{amsmath,upgreek,accents}
\definecolor{applegreen}{rgb}{0.55, 0.71, 0.0}

\definecolor{amber}{rgb}{1.0, 0.55, 0.1}

\usepackage{amssymb}

\usepackage{textgreek}
\newcommand{\beq}{\begin{eqnarray}}

\newcommand{\eeq}{\end{eqnarray}}

\newcommand{\bea}{\begin{eqnarray}}
\newcommand{\eea}{\end{eqnarray}}

\usepackage[T3,T1]{fontenc}

\usepackage[colorlinks,hyperindex,unicode]{hyperref}
\definecolor{green1}{RGB}{0,128,0} 
\hypersetup{hidelinks,backref=true,pagebackref=true,hyperindex=true,colorlinks=true,breaklinks=true,urlcolor= blue}
\hypersetup{%
  colorlinks = true,
  linkcolor  = blue,
  citecolor = cyan,
}
\usepackage{bookmark,textgreek}
\usepackage{hyperref,color,xcolor}

\newcommand\orcidroldao{{\href{https://orcid.org/0000-0003-3978-532X}{\orcidicon}}}
\newcommand\orcidkuntz{{\href{https://orcid.org/0000-0003-2621-5715}{\orcidicon}}}
\newcommand{\orcidicon}{%
	\begin{tikzpicture}
	\draw[lime, fill=lime] (0,0)
		circle [radius=0.16]
		node[white] {{\fontfamily{qag}\selectfont \tiny ID}};
	\draw[white, fill=white] (-0.0625,0.095)
		circle [radius=0.007];
	\end{tikzpicture}	\hspace{-2mm}
}

\def\nn{\nonumber }

\begin{document}
\title{Transport coefficients in AdS/CFT and quantum gravity corrections due to a functional measure}

\author{Iberê Kuntz
\orcidkuntz\!\!
}
\affiliation{Departamento de F\'isica, Universidade Federal do Paran\'a, PO Box 19044, Curitiba -- PR, 81531-980, Brazil.}
\email{kuntz@fisica.ufpr.br}

\author{Roldao da Rocha
\orcidroldao\!\!
}
\affiliation{Federal University of ABC, Center of Mathematics,  Santo Andr\'e, 09210-580, Brazil.}
\email{roldao.rocha@ufabc.edu.br}

\begin{abstract}
The presence of a functional measure is scrutinized on both sides of the dual  gauge/gravity correspondence.
Corrections to the transport coefficients in relativistic hydrodynamics are obtained using the linear response procedure. In particular, using first-order hydrodynamics, the shear viscosity, entropy density, diffusion constant, and speed of sound are shown not to acquire any corrections from the functional measure of gravity,   for a Minkowski background metric. On the other hand, the energy density, the pressure, the relaxation time, the bulk viscosity,  the decay rate of sound waves, and coefficients of conformal traceless tensor fields,  are shown to carry significant quantum corrections due to the functional measure, even for a flat background. They all acquire an imaginary part that reflects the instability of the strongly-coupled fluids on the boundary CFT.  This opens up the possibility of testing quantum gravity with the  quark-gluon plasma.
\end{abstract}
\maketitle
\section{Introduction}
\label{s1}
AdS/CFT consists of a holographic duality, leading a type IIB string theory with  effective compactification AdS${}_5 \times S^5$ to a   SU(${N}_c$) Yang--Mills theory with four supersymmetric charges, in the limit of a large number of colors $N_c$   \cite{malda}. Open strings with endpoints at a stack of ${N}_c$ coincident $D_3$-branes are dual to gauge fields with symmetry SU(${N}_c$). In the low-energy regime, the radius of the 5-sphere is negligible and the  bulk effectively becomes  AdS${}_5$.  
The lack of analytical and  perturbative solutions to strongly-coupled CFT can be circumvented by techniques in AdS/CFT, where 
strongly-coupled CFT can be investigated using weakly-coupled gravity on the bulk. Moreover, string theory
is more straightforward to tackle in the strong-coupling regime \cite{Nastase:2015wjb}.

In the long-wavelength regime, fluid/gravity duality sets in, where fluid flows can be 
characterized by transport coefficients. These coefficients take into account the microscopic hydrodynamical features  of fluid flows, which are based on interactions in the strong-coupled  field-theoretical aspects. The dual gravity in the bulk, represented by a black brane with a non-degenerate horizon, makes CFT, at long-length scales, to be regulated by the near-horizon domain of the bulk geometry \cite{Iqbal:2008by}.
The gauge/gravity correspondence is thus a useful tool to compute transport  coefficients \cite{Son:2007vk}.

In the context of the holographic  correspondence dictionary, operators describing gravitons in the AdS${}_5$ bulk are dual objects to the  energy-momentum  tensor, which regulates the strong-coupled CFT on the AdS${}_5$ boundary of codimension one. 
Taking into account the long-wavelength limit,  conservation of the  energy-momentum  tensor is led to hydrodynamics, and holography relates gravity in the bulk to fluid dynamics on the boundary. In the fluid/gravity correspondence scenario, the system of equations governing relativistic hydrodynamics on the boundary of AdS${}_5$ are directly associated with Einstein's equations in AdS${}_5$   \cite{Bhattacharyya:2007vjd,Haack:2008cp,Policastro:2002se,Bernardo:2018cow}.

One of the most pertinent  applications of AdS/CFT is the calculation of linearized hydrodynamics for a variety of gauge theories with holographic duals. 
Various features of the dual fluid dynamics on the AdS boundary, represented by the transport and response  coefficients, have been comprehensively studied \cite{Policastro:2001yc}, also including the relativistic Navier--Stokes equations with soft-hair, in the membrane paradigm \cite{Ferreira-Martins:2021cga,Ferreira-Martins:2019svk}. 
Fluid/gravity establishes how the gravitational dynamics in the AdS bulk, which is dictated by Einstein's field equations, relates to the hydrodynamics, which is described by the relativistic Navier--Stokes  equations in the AdS boundary. 
The latter are strongly hyperbolic, causal, and stable, also in the complete out-of-equilibrium regime proposed in the seminal BDNK  (Bemfica--Disconzi--Noronha--Kovtun) approach \cite{Bemfica:2020xym,Bemfica:2017wps,Rocha:2022ind,Kovtun:2019hdm,Hoult:2020eho}. Fermionic sectors of the gauge/gravity correspondence have been also scrutinized \cite{Meert:2018qzk,Bonora:2014dfa}.  Relativistic hydrodynamics plays a prominent role in phenomenological and theoretical approaches. In particular,  experiments at the Relativistic Heavy Ion Collider (RHIC) can be reasonably described by computational simulations regarding relativistic hydrodynamics \cite{DerradideSouza:2015kpt,Karapetyan:2018yhm}.

Among the most relevant  results underlying the fluid/gravity correspondence, the calculation of the shear viscosity-to-entropy density ratio, which is a natural observable, plays a prominent role  \cite{Kovtun:2004de}.  To calculate it, the literature usually adopts a semi-classical expansion, keeping only the classical contribution from the AdS bulk.
Ref. \cite{Kuntz:2019omq} pioneered the investigation of one-loop quantum gravitational corrections to the shear viscosity-to-entropy density ratio. It was shown that one-loop corrections to the shear viscosity vanish, but the entropy density gains a logarithmic quantum correction. The coefficients of the latter depend on the spin of the particles running in the loop, which violates the KSS (Kovtun--Son--Starinets) bound. Besides,  Ref. \cite{Montenegro:2020paq} formulated 
a causal theory with spin, proposing  corrections due to backreaction to hydrodynamical variables, within a one-loop effective action setup.

The goal of this work is to continue the study of quantum corrections in AdS/CFT. In particular, we shall investigate the contribution of the functional measure, which is itself a one-loop correction, to transport coefficients of the gauge theory.

In principle, there could be contributions from the functional measures of the gravitational and the gauge theory. These measures are dimensionless and independent of any physical parameters (such as Newton's constant $G$ or $N_c$). Therefore, the functional measures are not affected by gravitational nor gauge parameters. 
On the gravity side, the resulting effective action thus gains an additional (Planck-suppressed) correction:
\begin{align}
	\mathcal{L}_\text{eff} &= \frac{1}{16\pi G} R + \mathcal{L}_{measure}
	\\
	&=
	\frac{1}{16\pi G} \left( R + 16\pi G \mathcal{L}_{measure} \right)
	\ .
\end{align}
Using the AdS/CFT dictionary $N_c^2 \sim 1/G$, thus showing that the gravitational functional measure corresponds to a finite $N_c$ correction to the gauge theory. However, as we shall see, such corrections vanish. On the other hand, the functional measure on the gauge side yields non-vanishing corrections. We stress that the gauge functional measure is not obtained from mapping the gravitational measure into the gauge side via the duality. It rather results directly from the configuration-space geometry of the gauge theory and, therefore, is independent of AdS/CFT. The latter shall only play a role in the calculation of Green's functions.

The presence of a functional measure in the path integral is crucial for obtaining gauge-invariant results \cite{Vilkovisky:1984st,Mottola:1995sj,Fujikawa:1983im,Fujikawa:1979ay,Toms:1986sh,DeWitt:2003pm}. On the other hand, it requires the introduction of a metric in configuration space. This metric, as opposed to the gravitational metric, is not dynamical and, therefore, its choice must be seen as part of the definition of the theory. Indeed, the attempt of promoting the configuration-space metric to dynamical objects recovers gauge-dependence issues \cite{Casadio:2022ozp}.

We shall adopt symmetry as the main tool for building the configuration-space metric, in the same spirit as effective field theory. The most general functional measure, to the lowest order, can thus be obtained. The measure yields an additional term in the effective action, whose consequences for transport coefficients in the linear response theory shall be studied through a hydrodynamical representation of the energy-momentum tensor. 

This paper is organized as follows. In Sec. \ref{measure1}, we review some aspects of the functional measure in quantum field theory and the construction of the configuration-space metric, which ultimately results in an additional term in the effective action. 
Sec. \ref{s2} is dedicated to scrutinizing the functional measure on the gravity side, whereas Sec. \ref{s3} investigates the consequences of a  functional measure on the boundary CFT. After introducing the retarded Green’s response function  in the framework of perturbation theory, an  effective  energy-momentum is obtained, encoding additional quantum contribution  from
the functional measure. Perfect and viscous fluids are studied in this context. In particular, the energy density receives an imaginary contribution, thus suggesting the existence of instabilities in the fluid. The other transport coefficients also gain similar corrections, except for the shear viscosity. The shear viscosity-to-entropy density ratio is preserved. The second-order hydrodynamics expansion is also implemented in Sec. \ref{s4}, where corrections to the sound pole, to the relaxation time, and the effective diffusion constant are obtained, among other three transport coefficients that carry corrections due to the functional measure. Also, a reliable bound on the parameter that regulates quantum gravity corrections, due to the functional measure, is obtained by analyzing the latest experimental data about the bulk viscosity  of the quark-gluon plasma. 
Finally, conclusions, discussions, and perspectives are drawn in Sec. \ref{s5}.

\section{Functional measure in Quantum Field Theory}
\label{measure1}

A central object in quantum field theory is the generating functional:
\begin{equation}
	Z[J]
	=
	\int\mathrm{d}\mu[\varphi] e^{i \left( S[\varphi^i] + J_i \varphi^i \right)},
	\label{Z}
\end{equation}
where $S[\varphi]$ is the classical action for some generic set of fields $\varphi^i = (\phi(x), A_\mu(x), g_{\mu\nu}(x), \ldots)$. All information regarding any physical system is contained in its corresponding generating functional $Z[J]$. In particular, correlation functions are obtained by simple functional differentiations, from which scattering amplitudes can be obtained using the LSZ (Lehmann--Symanzik--Zimmermann) formula. Despite its importance in field theory, a rigorous mathematical foundation remains unknown, particularly with respect to the functional measure $\mathrm{d}\mu[\varphi]$. Operationally, one can however define the aforementioned object as
\begin{equation}
    \mathrm{d}\mu[\varphi] = \mathcal{D}\varphi^i \sqrt{\Det G_{ij}}
    \ ,
    \label{measure}
\end{equation}
where $\mathcal{D}\varphi^i = \prod_i \mathrm{d}\varphi^i$ and $\Det G_{ij}$ denotes the functional determinant of the configuration-space metric $G_{ij}$. The factor $\sqrt{\Det G_{ij}}$ is required to account for a non-trivial configuration space, whose typical example is that of a non-linear sigma model.

The presence of a non-trivial functional measure is usually sidestepped by employing dimensional regularization:
\begin{align}
    \Det G_{ij}
    &=
    e^{\delta^{(n)}(0) \int\mathrm{d}^nx \sqrt{-g} \, \tr\log G_{ij}}
    \nonumber
    \\
    &\stackrel{dim.\, reg.}{=} \ 1
    \ ,\label{dimr}
\end{align}
where we used $\Det\log = \Tr\log$. Here $\Tr$ and $\tr$ denote the functional and the matrix trace, respectively. The last equality in Eq. (\ref{dimr}) follows from formally setting $\delta^{(n)}(0) = 0$ in dimensional regularization. This argument is not strictly valid \cite{DeWitt:2003pm} and, therefore, it must be confirmed case-by-case with other regularization schemes. We shall later obtain non-trivial corrections to observables, namely transport coefficients. Our results thus serve as an \textit{a posteriori} proof that one cannot indeed appeal to dimensional regularization to set $\delta^{(n)}(0) = 0$ in this case.

We shall regularize the Dirac delta using the Gaussian distribution:
\begin{align}
    \delta^{(n)}(x)
    &=
    \frac{1}{(2\pi)^{n/2} \ell^n} e^{\frac{-x^2}{2 \ell^2}}
    \ ,
\end{align}
for some (length) cutoff $\ell \to 0$. When evaluated at the origin, we thus find
\begin{equation}
    \delta^{(n)}(0) = \frac{1}{(2\pi)^{n/2} \ell^n}
    \ .
    \label{delta}
\end{equation}
After renormalization, Eq.~\eqref{delta} implies a non-trivial contribution from the configuration-space metric in \eqref{measure} to the path integral. It is the phenomenological consequences of this contribution to transport coefficients that we are interested in this article.

It should be stressed that there is no canonical way of choosing the configuration-space metric, which must be seen as part of the definition of the theory. The configuration-space metric is typically identified with the bilinear form appearing in the kinetic term \cite{Meetz:1969as,Slavnov:1971mz,Vilkovisky:1984st}, in which case it can be chosen to cancel ultralocal loop divergences \cite{Fradkin:1973wke,Fradkin:1976xa}. We stress, however, that this identification is not mandatory. There is no physical argument to justify such an ad-hoc procedure. For this reason, we shall construct the configuration-space metric based on symmetry principles alone, in the same spirit as effective field theories. In this case, the ultralocal divergences are not automatically canceled and, therefore, must be renormalized.

Sticking to the lowest order in the energy-scale expansion, which does not include dimensionful parameters, and assuming ultralocality, yields \cite{Casadio:2021rwj,Kuntz:2022tat}~\footnote{The geometry of configuration space was also studied in connection with the problem of singularities in Ref.~\cite{Casadio:2020zmn}.}
\begin{align}
G_{ij}
=
{\rm diag}\,
\Bigg(
\overbrace{G_{IJ}^{\rm s}, \ldots, G_{IJ}^{\rm s}}^{N_{\rm s}},
\overbrace{G_{IJ}^{\rm f}, \ldots, G_{IJ}^{\rm f}}^{N_{\rm f}},
&\overbrace{G_{IJ}^{\rm YM},\ldots, G_{IJ}^{\rm YM}}^{N_{\rm YM}},
\nonumber
\\
&\overbrace{G_{IJ}^{\rm nYM},\ldots, G_{IJ}^{\rm nYM}}^{N_{\rm nYM}},
G_{IJ}^{\rm DW}
\Bigg) \, \delta(x,x')
\ ,
\label{fullmetric}
\end{align}
where
\begin{align}
    G_{IJ}^{\rm s}
    &=
    \sqrt{-g}
    \ ,
    \\
    G_{IJ}^{\rm f}
    &=
    \sqrt{-g}
    \left(
        \begin{matrix}
        0 & \varepsilon_{\alpha\beta} \\
        \varepsilon_{\alpha\beta} & 0
    \end{matrix}
    \right)
    \ ,
    \\
    G_{IJ}^{\rm YM}
    &=
    \sqrt{-g}\, g_{\mu\nu}
    \ ,
    \\
    G_{IJ}^{\rm nYM}
    &=
    \sqrt{-g}\, g_{\mu\nu}\, \delta_{ab}
    \ ,
    \\
    G_{IJ}^{\rm DW}
	&=
	\frac{\sqrt{-g}}{2}
	\left(
	g_{\mu\rho}\, g_{\sigma\nu}
	+ g_{\mu\sigma}\, g_{\rho\nu}
	+ c\, g_{\mu\nu}\, g_{\rho\sigma}
	\right)
	\ ,
\end{align}
denote the configuration-space metric for scalars, fermions, abelian Yang--Mills fields, non-abelian Yang--Mills fields, and gravity, respectively. Here $\varepsilon_{\alpha\beta}$ denotes the inverse of the two-dimensional Levi--Civita tensor, $\delta_{ab}$ is the metric in gauge space and $c$ is a dimensionless free parameter. Note that the gravitational sector is described by the well-known DeWitt metric $G_{IJ}^{\rm DW}$.
The functional determinant, in this case, can be readily calculated to wit:
\begin{equation}
	\Det G_{ij}
	=
	e^{2\, \tilde\gamma \int\mathrm{d}^n x \sqrt{-g} \, \tr\log |g_{\mu\nu}|}
	\ ,
	\label{detG2}
\end{equation}
where
\begin{align}
    \tilde\gamma(\ell) &= \frac{\beta}{2 (2\pi)^{n/2} \ell^n} \ ,
    \\
    \beta
    &=
    \frac{1}{2}\,N_{\rm s}
    +
    D\,N_{\rm f}
    +
    \frac{(n - 2)}{2}\,N_{\rm YM}
    +
    \frac{(n - 2)}{2} \left(N_g^2 - 1\right)
    N_{\rm nYM}
    +
    \frac{1}{4} (n - 4)(n + 1)
    \ .
\end{align}
We have also omitted a divergent constant factor, which is otherwise canceled by the path integral normalization $Z[0] = 1$.

Eqs.~\eqref{Z} and \eqref{detG2} then yields the generating functional in the more standard form
\begin{equation}
	Z[J]
	=
	\int \mathcal{D}\varphi^i e^{i \left( S_\text{eff}[\varphi^i] + J_i \varphi^i \right)},
\end{equation}
with a renormalized effective action
\begin{equation}
    S_\text{eff} = \int\mathrm{d}^nx \sqrt{-g}
    \left(
    \mathcal L
	-
    i \gamma \, \tr\log |g_{\mu\nu}|
	\right)
	\ ,
	\label{eq:newac}
\end{equation}
for some bare Lagrangian $\mathcal L$. Here $\gamma$ is the renormalized parameter, whose (Wilsonian) renormalization group equation reads
\begin{equation}
    \Lambda \frac{d\gamma}{d\Lambda} = -\frac{n \beta}{2 (2\pi)^{n/2}} \Lambda^n
    \ ,
    \label{RGE}
\end{equation}
where $\Lambda = 1/\ell$.

A few comments are in order. When $\hbar$ is reinstated, it is straightforward to see that the contribution from the measure in Eq.~\eqref{eq:newac} is of one-loop order $\mathcal O(\hbar)$. The usual one-loop correction $\Tr\log(S_{,ij})$ should thus be taken into account in this order. The latter, however, entails a series in powers of curvatures and/or covariant derivatives, which are subdominant at low energies when compared to the derivative-free measure contribution. We shall thus focus on the one-loop correction due to the functional measure. The effects of $\Tr\log(S_{,ij})$ have been considered in \cite{Kuntz:2019omq}.

Furthermore, note that there is an apparent violation of diffeomorphism invariance in Eq.~\eqref{eq:newac}. This is indeed expected because the functional measure \eqref{measure} transforms as a (functional) scalar density in such a way that the whole path integral is invariant under reparameterizations. However, we stress that variations of the symmetry-breaking term under spacetime diffeomorphisms are canceled by the functional Jacobian that shows up from $\mathcal{D}\varphi^i$, thus the path integral and all observables in the quantum regime remain invariant.

\section{Functional measure on the gravity side}
\label{s2}
On the gravity side of the GKP-Witten relation, \cite{Witten:1998qj,Gubser:1998bc}
\begin{equation}
    Z_\text{gauge}
    =
    Z_\text{gravity}
    \ ,
\end{equation}
the correction in Eq.~\eqref{eq:newac} modifies the Einstein-Hilbert action, thereby altering the dynamics of the bulk field $h_{\mu\nu}$. In this section, we shall solve the modified equations of motion for $h_{xy}$. In particular, the on-shell action $S[h_{xy}]$ acquires no additional finite term.

The modified action reads
\begin{equation}
    S
    =
    S_{EH}
    + S_{\mathcal M}
    \ ,
\end{equation}
where $S_{EH}$ is the Einstein-Hilbert action with negative cosmological constant and $S_{\mathcal M}$ denotes the new term due to the functional measure. We split the metric into some (asymptotically-AdS) background $\bar g_{\mu\nu}$ and perturbations $h_{\mu\nu}$, namely
\begin{equation}
    g_{\mu\nu}
    =
    \bar g_{\mu\nu}
    + h_{\mu\nu}
    \ ,
\end{equation}
to wit
\begin{align}
    S_{\mathcal M}
    =
    -i\gamma
    \int\mathrm{d}^nx \sqrt{-\bar g}
    \bigg[ &
        \tr\log |\bar g_{\mu\nu}|
        + \left(1 + \frac12 \tr\log|\bar g_{\mu\nu}|\right)
        \left( h - \frac12 h^\rho_{\ \nu} h^\nu_{\ \rho} \right)
        \nonumber\\
        & + \frac12 \left(
            1
            + \frac14 \tr\log|\bar g_{\mu\nu}|
            \right) h^2
    \bigg]
    \ ,
\end{align}
where $h = \bar g^{\mu\nu} h_{\mu\nu}$. The corresponding corrections to the equations of motion then read
\begin{equation}
    -i \gamma \left(
        1
        + \frac12 \tr\log |\bar g_{\rho\sigma}|
        \right) \bar g_{\mu\nu}
    +i \gamma \left(
        1
        + \frac12 \tr\log |\bar g_{\rho\sigma}|
    \right) h_{\mu\nu}
    - i \gamma \left(
        1
        + \frac14 \tr\log |\bar g_{\rho\sigma}|
        \right) h \bar g_{\mu\nu}
    \ .
\end{equation}
For a diagonal background metric, such as the  Schwarzschild-AdS one, the dynamics of $\varphi \equiv h^x_{\ y}$ becomes:
\begin{equation}
    \Box\varphi
    + i \gamma \left(
        1
        + \frac12 \tr\log |\bar g_{\rho\sigma}|
        \right) \bar g_{xx} \varphi
    = 0
    \ ,
    \label{eq:neweom}
\end{equation}
where $\Box = \bar g^{\mu\nu} \nabla_\mu\nabla_\nu$ and covariant derivatives are defined with the background metric $\bar g_{\mu\nu}$.

We now take the Schwarzschild-AdS background~\footnote{We have conveniently set all dimensionful parameters to unity.}
\begin{equation}
    ds^2
    =
    \frac{1}{u^2}
    \left(
        -h(u) dt^2
        + d\vec{x}^2
    \right)
    + \frac{1}{h(u) u^2} du^2
    \ ,
\end{equation}
with $h(u) = 1 - u^4$. Assuming, for simplicity, that $\varphi = \varphi(u)$, Eq.~\eqref{eq:neweom} yields
\begin{equation}
    u^2 \varphi''
    - 3 u \varphi'
    - 5 i \gamma \frac{\log u}{u} \varphi
    = 0
    \label{eq:eom2}
\end{equation}
in the asymptotic region ($u\to 0$).

Note that the branch point singularity in $\log u$ prevents us from finding a typical solution in a power series of $u$. We shall thus solve Eq.~\eqref{eq:eom2} perturbatively in $\gamma$. For $\gamma = 0$, the asymptotic solution has the well-known form:
\begin{equation}
    \varphi_\text{sol}(u)
    =
    \varphi^{(0)} (1 + \varphi^{(1)} u^4)
    \ .
    \label{eq:sol}
\end{equation}
For $\gamma \neq 0$, we shall write
\begin{equation}
    \varphi(u)
    =
    \varphi_\text{sol}(u)
    -i \gamma \, \xi(u)
    \ ,
    \label{eq:pertg}
\end{equation}
where $\xi(u)$ is an arbitrary function whose functional form is to be found. Using Eq.~\eqref{eq:pertg} in Eq.~\eqref{eq:eom2} and keeping only leading order terms in $\gamma$, we obtain
\begin{equation}
    u^2 \xi'' - 3u \xi' + 5 \varphi^{(0)} \frac{\log u}{u} = 0
    \ ,
\end{equation}
whose most general (exact) solution is
\begin{equation}
    \xi(u)
    =
    c_1 + c_2 u^4
    - \frac{6 \varphi^{(0)}}{5 u}
    - \frac{\log u}{u} \varphi^{(0)}
    \ .
\end{equation}
The first two terms have the same form as Eq.~\eqref{eq:sol}, thus they can be absorbed by a finite renormalization of $\varphi^{(0)}$ and $\varphi^{(1)}$. In particular, they do not affect correlation functions in the gauge theory because the latter are obtained from variations with respect to $\varphi^{(0)}$. The last two terms, on the other hand, are divergent at the AdS boundary and must be eliminated by holographic renormalization. Indeed, the on-shell action reads
\begin{equation}
    S
    =
    \int\mathrm{d}^4x
    \left[
        2 {\varphi^{(0)}}^2 \varphi^{(1)}
        + 2 i \gamma {\varphi^{(0)}}^2 \varphi^{(1)} \frac{\log u}{u}
        + i \gamma \frac{{\varphi^{(0)}}^2}{2 u^6}
        + \mathcal{O}(\gamma^2)
    \right]
    \quad\quad (u\to 0)
    \ .
\end{equation}
After canceling all divergences, no non-trivial corrections are thus found for the correlation functions of $T_{xy}$ in the gauge theory due to the functional measure in $Z_\text{gravity}$. The same argument can be straightforwardly generalized to the other components of $h_{\mu\nu}$, whose equation of motion also exhibits branch-cut divergences that yield no corrections to correlation functions of $T_{\mu\nu}$ on the gauge theory after holographic renormalization.

\section{Functional measure on the gauge side: linear response theory}
\label{s3}
The linear response theory can be used to investigate the  response of a quantum-mechanical system to perturbation theory. The Hamiltonian can be written as $
H=H_0+\delta H(t),$ 
where $H_0$ is the free Hamiltonian and the perturbed Hamiltonian and action are respectively regarded as \cite{Natsuume:2014sfa}
\begin{align}
\delta H(t) &= - \int d^3x\, \mathcal{O}(\bmx) \source(t,\bmx)~, \\
\delta \action &= \int d^4x\,  \mathcal{O}(\bmx)\source(t,\bmx)~,
\label{eq:perturb}
\end{align}
where  the external source $\source$ carries the response to the operator $\mathcal{O}$, whose  average under small  perturbations reads 
\beq
\langle \mathcal{O}(t,\bmx) \rangle_{{\rm av}} = \text{tr} \left[ \uprho(t) \mathcal{O}(\bmx) \right],
\label{123}
\eeq
for the subscript on the left-hand side of Eq. (\ref{123}) representing the average in the presence of the external source and  $\uprho(t_0)$ denoting the  density matrix associated with a canonical ensemble. Turning on the source  $\source$ at $t=t_0$, the density matrix evolves, 
yielding 
\beq
\langle \mathcal{O}(t,\bmx) \rangle_{\rm av}
= \text{tr} \left[ \uprho(t_0) U^{-1}(t,t_0) \mathcal{O}(\bmx) U(t,t_0) \right].
\label{eq:evol_Heisenberg}
%
\eeq
The operator in the interaction picture reads $\mathcal{O}_I = e^{iH_0(t-t_0)} \mathcal{O} e^{-iH_0(t-t_0)}$. From the Schr\"{o}dinger equation, considering  
%
$\delta H_I := e^{iH_0(t-t_0)} \delta H e^{-iH_0(t-t_0)}$, 
%
and keeping only first-order terms in $\source$, yields
\begin{align}
\langle \mathcal{O}(t, \bmx) \rangle_{\rm av} 
&= \text{tr} \left[ \uprho(t_0) \mathcal{O}_I(t, \bmx) \right]  - i\, \text{tr} \left[ \uprho(t_0) \int_{t_0}^t d\mathsf{t} 
[\mathcal{O}_I(t,\bmx), \delta H_I(\mathsf{t}) ] \right] + \cdots.
\end{align}
%
%
Taking the $t_0 \rightarrow -\infty$ limit therefore implies that 
\begin{align}
\delta \langle \mathcal{O}(t, \bmx) \rangle 
&:= \langle \mathcal{O}(t, \bmx) \rangle_{\rm av} - \langle \mathcal{O} \rangle=  
- i\int_{\mathbb{R}} d^4x'\, 
G_R^{\mathcal{O}\mathcal{O}}(t-\mathsf{t}, \bmx-\bmx') \source(\mathsf{t},\bmx'),
\label{eq:linear_response}
\end{align}
where the retarded Green's response function reads
\beq
G_R^{\mathcal{O}\mathcal{O}}(t-\mathsf{t}, \bmx-\bmx') := -i \theta(t-\mathsf{t})
\left\langle [\mathcal{O}(t,\bmx), \mathcal{O}(\mathsf{t},\bmx') ] \right\rangle
\label{grr}
\eeq
and $\theta(t-\mathsf{t})$ denotes the Heaviside step function.
The Fourier transform of Eq. (\ref{grr}) yields
\be
{
\delta \langle \mathcal{O}(k) \rangle = - \,G_R^{\mathcal{O}\mathcal{O}}(k)\source(k)~,
}
\label{eq:linear_response:mom}
\ee
where $q_\mu = (\omega, \bmq)$ and the response function reads \be
G_R^{\mathcal{O}\mathcal{O}}(k) = -i \int_{\mathbb{R}} d^4x\, e^\ikx 
\theta(t)  \left\langle [\mathcal{O}(t,\bmx), \mathcal{O}(0,\vec{0}) ] \right\rangle,
\ee
which can be computed using AdS/CFT. 
In fluid dynamics, a perturbed Lagrangian is given by
\beq
    \delta{\cal L} = h^{(0)}_{\mu\nu}(t) T^{\mu\nu}(\vec{x}),
\eeq
and 
\beq
    \delta \langle T^{\mu\nu} \rangle &=& -G_R^{\mu\nu,\rho\sigma} h^{(0)}_{\rho\sigma}(t), 
    \label{eq:response_emtensor} \\
    G_R^{\mu\nu,\rho\sigma} &=& -i \int_{\mathbb{R}} d^4x\, e^\ikx 
    \theta(t)  \left\langle [T^{\mu\nu}(t,\bmx), T^{\rho\sigma}(0,\vec{0}) ] \right\rangle~.\label{rgf}
\eeq
The energy-momentum tensor couples to metric perturbations at the boundary and, for the action \eqref{eq:newac}, it is given by
\begin{align}
    T_{\mu\nu}^\text{eff}
    &=
    \frac{2}{\sqrt{-g^{(0)}}}
    \left[
        \frac{\partial(\sqrt{-g^{(0)}} \mathcal L_\text{eff})}{\partial {g^{(0)}}^{\mu\nu}}
        - \partial_\alpha \frac{\partial(\sqrt{-g^{(0)}} \mathcal L_\text{eff})}{\partial(\partial_\alpha {g^{(0)}}^{\mu\nu})}
    \right]\nonumber
    \\
    &=
    T_{\mu\nu}
    + 2 i \gamma \left(1 + \frac12 \tr\log |g_{\rho\sigma}^{(0)}| \right) \, g_{\mu\nu}^{(0)}
    \ ,
    \label{eq:effT}
\end{align}
where $T_{\mu\nu}$ corresponds to the classical contribution from $\mathcal L$ and the last term is the contribution from the functional measure. The background metric $g_{\mu\nu}^{(0)}$ has been kept arbitrary so far. In the following, we shall restrict our results to the flat background, which is the most important case for applications in the quark-gluon plasma (QGP), as a byproduct of  heavy-ion collisions described by a strongly-coupled fluid that
can be identified by its underlying  macroscopic features, including the equation of state and transport and response coefficients. The microscopic properties also involve interactions among the fluid constituents, dictated by QCD.  We stress, nonetheless, that curved backgrounds on the gauge side of the correspondence also play important roles in other contexts \cite{Casadio:2016zhu}.

\subsection{Perfect fluid flow}

Taking into account fluids that have no particle number conserved, just the energy-momentum tensor can be associated with a locally-conserved current. In this scenario, fluctuations of  momentum densities and energy represent the only hydrodynamic modes \cite{Natsuume:2014sfa}.
Fluids are thus solely described by their energy-momentum tensor, which is covariantly conserved. In particular, notice that the quantum contribution in $T^\text{eff}$ only depends on the metric, thus metric compatibility gives
\begin{equation}
    \nabla^\mu T^\text{eff}_{\mu\nu}
    =
    0
    \ ,
\end{equation}
which is indeed expected from the diffeomorphism invariance of the path integral.

In a perfect fluid description, at least as a first approximation, the rate of momentum transport, encoded into  the shear and bulk viscosities, can be neglected. Its constitutive equation arises from the $0^{\rm th}$-order terms in the derivative expansion of the energy-momentum tensor:
\be
\left(T^\text{eff}\right)^{\mu\nu} = \left(\upepsilon^\text{eff}+P^\text{eff}\right)u^\mu u^\nu + P^\text{eff} {g^{(0)}}^{\mu\nu},
\label{eq:EM_constitutive}
\ee
where $u^\mu(x)$ is the four-velocity field of the fluid flow, $P^\text{eff} = (T^\text{eff})^{11}$ denotes the effective pressure, and $\upepsilon^\text{eff} = (T^\text{eff})^{00}$ is the energy density in the (local) {rest frame}.
Comparing Eqs.~\eqref{eq:effT} and \eqref{eq:EM_constitutive}, gives
\begin{align}
    \upepsilon^\text{eff}
    &=
    \upepsilon
    - 2 i \gamma
    \ ,
    \label{em2}
    \\
    P^\text{eff}
    &=
    P + 2i\gamma
    \ ,
    \label{peff}
\end{align}
where we have set $g^{(0)}_{\mu\nu} = \eta_{\mu\nu}$ as the flat metric. Here $\upepsilon = T_{00}$ denotes the classical energy density and $P$ is the classical pressure. The imaginary part of the energy density \eqref{em2} corresponds to the instability of the fluid's degrees of freedom. In particular, $\Im(\upepsilon^\text{eff})$ is a measure of the fluid flow lifetime. This is, in fact, consistent with the difficulty of creating a long-lasting QGP in the laboratory. The strength of this instability is measured by the free parameter $\gamma$, which must be fixed by experiments. We will address bounds on $\gamma$ analyzing the latest experimental results about the QGP.

\subsection{Viscous fluid}
For viscous fluids, the constitutive equation must be generalized to include first-order derivatives of the fluid flow velocity:
\begin{align}
(T^\text{eff})^{\mu\nu} &= \left(\upepsilon^\text{eff}+P^\text{eff}\right)u^\mu u^\nu + P^\text{eff} g^{(0)\mu\nu} 
+ (\uptau^\text{eff})^{\mu\nu}~, 
\label{eq:constitutive_curved} 
\\
(\uptau^\text{eff})^{\mu\nu} &= 
- \Upupsilon^{\mu\alpha} \Upupsilon^{\nu\beta} \left[
\upeta^\text{eff} \left(\nabla_{(\alpha} u_{\beta)}  
- \frac{2}{3} g^{(0)}_{\alpha\beta} \nabla \cdot  \vec{u} \right) 
+ \zeta^\text{eff} g^{(0)}_{\alpha\beta} \nabla \cdot  \vec{u} 
\right]~,
\label{eq:dissipative_curved}
\end{align}
where 
 $\Upupsilon^{\mu\nu} := g^{(0)\mu\nu} + u^\mu u^\nu$ is a projection operator along spatial directions. The effective shear and bulk viscosities are respectively denoted by $\upeta^\text{eff}$ and $\zeta^\text{eff}$.
 The shear viscosity naturally arises as the response under an internal thermal force, and specifies the relaxation of transverse momentum density fluctuations, whereas  the effective bulk viscosity measures how much the fluid is dislocated out of equilibrium in a uniform expansion \cite{Torrieri:2007fb}. Transport coefficients encode the mean free path (mfp) of any scattering process that is 
responsible for relaxation phenomena related to hydrodynamic modes \cite{Finazzo:2014cna}.  In a  conformal theory, the bulk viscosity vanishes, whereas, in a non-conformal relativistic theory, it is proportional to the mfp for processes lacking conservation of particle
number. It indeed consists of a uniform expansion, since (re-)achieving the equilibrium at a different temperature requires the change of the total number of particles. These sought-after viscosities are the crucial difference with respect to the perfect fluid and they are both entailed by the tensor $\uptau^\text{eff}$.

The shear viscosity, in particular, can thus be obtained by the response of the energy-momentum tensor to a small perturbation of the metric, playing the role of the strain. As the energy-momentum tensor consists of the variation of the Lagrangian density with respect to the metric, the underlying response has first-order in the derivative expansion, of the energy-momentum to the strain field. If one regards the dissipative contributions in the energy-momentum tensor,  the associated conservation equations correspond to the 
relativistic Navier--Stokes  equations \cite{Finazzo:2014cna}.

To find the viscosity, we perform a metric perturbation of the form:
\be
g^{(0)}_{\mu\nu} = 
\begin{pmatrix}
  -1 & 0 & 0 & 0 \\
  0 & 1 & h^{(0)}_{xy}(t) & 0 \\
  0 & h^{(0)}_{xy}(t) & 1 & 0 \\
  0 & 0 & 0 & 1 
\end{pmatrix}~.
\label{eq:metric_kubo}
\ee
Using \eqref{eq:dissipative_curved} one can calculate $\uptau^{xy}$ to linear order in the perturbation, as 
\be
\nabla_x u_y 
= \del_x u_y - \Gamma^\rho_{xy} u_\rho
= \frac{1}{2} \del_t h^{(0)}_{xy},
\ee
being $(\nabla \cdot \vec{u})$ second order in the perturbation. Hence Eq. \eqref{eq:dissipative_curved} reads 
\be
\delta \langle \uptau^{xy}  \rangle 
= 
 -2 \upeta^\text{eff} \Gamma^0_{xy}  
= - \upeta^\text{eff} \del_0 h^{(0)}_{xy},
\ee
whose Fourier transform is given by 
\be
{
\lim_{q\to0}\delta \langle \uptau^{xy}(\omega,\bmq)  \rangle = i\omega \upeta^\text{eff} \sourceG~.
}
\label{eq:Txy_vs_eta}
\ee
Comparing \eqref{eq:response_emtensor} and \eqref{eq:Txy_vs_eta} gives the Kubo formula, expressing $\upeta^\text{eff}$ as a function of the imaginary part of the retarded Green's response   function \eqref{rgf}:
\be
\upeta^\text{eff} = - \lim_{\substack{\omega\rightarrow 0\\
 q\to0}} \frac{1}{\omega} \Im\, G_R^{xy,xy}(\omega,\bmq)~.
\label{eq:viscosity_kubo}
\ee
Analogously, the retarded Green's response function in Eq. (\ref{rgf}) gives the bulk viscosity \cite{Gubser:2008sz,Buchel:2007mf}:
\beq
\zeta^{\rm eff}=-\frac{4}{9}\lim_{\substack{\omega\rightarrow 0\\
 q\to0}} \frac{1}{\omega}\Im G_R^{\mu\mu,\rho\rho}(\omega,\bmq).\label{rgf1}
\eeq
Implementing the perturbation \eqref{eq:metric_kubo} in Eq.~\eqref{eq:effT} results in
\begin{equation}
    T^\text{eff}_{\mu\nu}
    =
    T_{\mu\nu}
    + 2i\gamma \eta_{\mu\nu}
    \ .
    \label{finalT}
\end{equation}
Comparing Eq.~\eqref{finalT} with Eqs.~\eqref{eq:viscosity_kubo} and \eqref{rgf1} gives
\begin{align}
    \upeta^\text{eff}
    &=
    \upeta
    \\
    \zeta^{\rm eff}
    &=\zeta+4i\gamma\,.\label{bul}
\end{align}
We, therefore, see that the shear viscosity does not receive any corrections from the functional measure for a Minkowski background, whereas the bulk viscosity does present a significant quantum correction. For the shear viscosity, Eq.~\eqref{eq:effT} indeed shows that no corrections exist for diagonal metrics.
Only non-diagonal (curved) background metrics can provide corrections to the shear viscosity. The components of a curved metric are nonetheless expected to be negligible at the scales of the laboratory.

It is also worth mentioning that the leading-order expression for the entropy production rate can be obtained
 from the energy-momentum conservation  
combined with the definition of the entropy density near equilibrium \cite{Buchel:2016cbj}:
\beq
s^{\rm eff} = \frac{\upepsilon^{\rm eff}+P^{\rm eff}}{T} = s,
\eeq
which recovers the classical entropy because the quantum corrections in Eqs.~\eqref{em2} and \eqref{peff} cancel out in the sum.
Therefore, the effective shear viscosity-to-entropy  density ratio does not receive any quantum correction from the functional measure, in first-order hydrodynamics:
\beq
\frac{\upeta^{\rm eff}}{s^{\rm eff}}
=
\frac{\upeta}{s}.
\eeq

\section{Second-order hydrodynamics and conformal fluids}
\label{s4}

Although first-order hydrodynamics cannot probe any functional measure signature in transport and response coefficients,
in what follows we will show that transport coefficients in the second-order hydrodynamics carry quantum gravity signatures of the functional measure.
 One can express all the possibilities regarding second-order terms in the energy-momentum tensor, permitted by Weyl
invariance \cite{Finazzo:2014cna}. 
 Then, one can compute the coefficients accompanying these
terms in the ${\cal N}=4$ SYM dual plasma \cite{Arnold:2011ja,Kovtun:2011np}.
In the context of  kinetic theory, the shear and bulk viscosities correspond to their respective mfps $\ell_\upeta$ and $\ell_\zeta$, which represent    scales at a microscopic level. As the effective energy density,  $\upepsilon^{\rm eff}$, and the fluid flow velocity, $u^\mu$, are assumed to be slowly-varying spacetime-dependent functions, a
macroscopic length scale $L$ can be therefore associated with their gradients, in  such a way that $\ell_\upeta,\ell_\zeta\ll L$. Therefore, any expansion term has order $\sim1$ in the respective small Knudsen numbers ($K_\zeta = \ell_\zeta/L$ and $K_\upeta = \ell_\upeta/L$  \cite{Finazzo:2014cna}) terms (e.g. $\upeta\upsigma_{\mu\nu}\sim 1$ in Eq. (\ref{gr1})). A continuous hydrodynamical description of the system as a fluid flow is dependent on assuming that $K_\zeta$ and $K_\upeta$ are small. Nonetheless, since dissipative terms solely emerge at order $\sim 1$ in this expansion, one may besides assume that the fluid flow is such that higher-order terms can be taken into account. Eq. (\ref{gr1}) suggests that  terms due to higher-order Knudsen numbers corrections are necessary, encoding dissipative components of non-conformal relativistic hydrodynamics. Formal aspects were developed in Ref. \cite{Montenegro:2016gjq}. 
 In this way the fluid flow admits higher-order
terms in the energy-momentum tensor.
Second-order hydrodynamics techniques  have been employed in the early stages of nuclear collisions, in the ultrarelativistic regime  \cite{Schenke:2012wb}, where the effective energy density has considerable gradients. It corresponds to second-order corrections in $K_\zeta$ and $K_\upeta$, which become relevant to the analysis of the QGP formation \cite{GoncalvesdaSilva:2017bvk,Oliveira:2020yac,Braga:2022yfe}. Hence, a second-order expansion is a necessary tool for applications \cite{Romatschke:2009kr}, in particular to relevant aspects of generalized black branes \cite{Casadio:2016zhu}.

The dissipative part, $(\Uppsi^\text{eff})^{\mu\nu}$, of the energy-momentum tensor
\begin{equation}\label{eq:Pimunu-def}
  (T^\text{eff})^{\mu\nu} = (\upepsilon^\text{eff} + P) u^\mu u^\nu + P^\text{eff} \Upupsilon^{\mu\nu} + (\Uppsi^\text{eff})^{\mu\nu}\,,
\end{equation}
 considers only the derivatives and vanishes
in a homogeneous equilibrium state. The tensor $(\Uppsi^\text{eff})^{\mu\nu}$ is
symmetric, transverse, and, for
conformal fluids, it is also  traceless. One defines 
\begin{equation}\label{sigmamunu}
  \upsigma^{\mu\nu} = 2{}^\llcorner\nabla^{\mu} u^{\nu}{}^\lrcorner\, , 
\end{equation}
where the notation 
\begin{equation}
  {}^\llcorner B^{\mu\nu}{}^\lrcorner
 = \frac12 \Upupsilon^{\mu\alpha} \Upupsilon^{\nu\beta}
     B_{(\alpha\beta)} 
  - \frac1{3} \Upupsilon^{\mu\nu} \Upupsilon^{\alpha\beta} B_{\alpha\beta}\, 
\equiv B^{\llcorner\mu\nu\lrcorner}
\end{equation} 
is employed for some given second rank tensor $B^{\mu\nu}$. Ref. \cite{Baier:2007ix} showed that the dissipative part of the
 energy-momentum  tensor, up to second order in derivatives, involves the Riemann and Ricci tensor, namely
\beq\label{gr1}
(\Uppsi^\text{eff})^{\mu\nu} &=& -\upeta \upsigma^{\mu\nu}- \upeta\tau_\Uppi \left[  u_\rho\nabla^\rho {}^\llcorner\upsigma^{\mu\nu}{}^\lrcorner 
 + \frac 12 \upsigma^{\mu\nu}
    (\nabla \cdot \vec{u}) \right]  
  + \upkappa\left[R^{\llcorner\mu\nu\lrcorner}-2 u_\alpha R^{\alpha\llcorner\mu\nu\lrcorner\beta} 
      u_\beta\right]\nonumber\\
  &&+ \frac{\uplambda_1}{\upeta^2} {\upsigma^{\llcorner\mu}}_\lambda \upsigma^{\nu\lrcorner\lambda}
  - \frac{\uplambda_2}{\upeta} {(\Uppsi^\text{eff})^{\llcorner\mu}}_\lambda \Upomega^{\nu\lrcorner\lambda}
  + \uplambda_3 {\Upomega^{\llcorner\mu}}_\lambda \Upomega^{\nu\lrcorner\lambda}\,,
\eeq
where  the vorticity can be expressed by 
\begin{equation}
  \Upomega^{\mu\nu} = 
  \frac12   
    \Upupsilon^{\mu\alpha}\Upupsilon^{\nu\beta}
    \nabla_{[\alpha} u_{\beta]}\,.
\end{equation}
The coefficient 
\beq\label{kk}
\upkappa =\lim_{\substack{{q}\to 0\\\omega\to 0}}\frac{\partial^2}{\partial{q^2}} G^{xy,xy}_R(\omega,\vec{q})
\eeq 
contributes to the 2-point Green's function of the
 energy-momentum  tensor  \cite{Baier:2007ix,Finazzo:2014cna}, whereas  the relaxation time for the shear viscous
stress, \beq
 \tau_\Uppi=\frac{1}{2\upeta}\left(\lim_{\substack{{q}\to 0\\\omega\to 0}}\frac{\partial^2}{\partial{\omega^2}} G^{xy,xy}_R(\omega,\vec{q})-\upkappa+T\frac{d\upkappa}{dT}\right),\label{tt}
\eeq   emerges both in the shear and sound modes \cite{Baier:2007ix,Finazzo:2014cna,Natsuume:2007ty,Rocha:2022fqz,Rocha:2021zcw}. 
The  $\uplambda_1$, $\uplambda_2$, and $\uplambda_3$ are second-order transport coefficients of conformal traceless tensor fields that are orthogonal to $u^\mu$ \cite{Romatschke:2009kr} and can be obtained,  
if one generalizes Eq. (\ref{eq:response_emtensor}), to encompass second-order hydrodynamics.  The energy-momentum tensor can be expressed as a Taylor-like expansion in the metric perturbation, as \cite{Arnold:2011ja}
\bea
\langle T^{\mu\nu}(\vec{x})\rangle_h&=&\langle T^{\mu\nu}\rangle_{h=0}-\frac 12
\int d^4 \vec{x}^\prime \,G^{\mu\nu, \sigma\rho}_R({x};\vec{x}^\prime)h_{\sigma\rho}(\vec{x}^\prime)
\nn\\
&&+
\frac 18\int d^4 \vec{x}^\prime\!\int d^4 \mathtt{x}\, G^{\mu\nu,\sigma\rho, \tau\xi}_{R}(\vec{x};\vec{x}^\prime,\mathtt{x})
h_{\sigma\rho}(\vec{x}^\prime) h_{\tau\xi}(\mathtt{x})+\dots\label{npo}
\eea
where $G^{\mu\nu,\dots}_R$ are retarded $n$-point correlators. By identifying the terms of the solution of the energy-momentum conservation law to the respective terms of the expansion in retarded correlators (\ref{npo}), hydrodynamic transport  coefficients can be expressed by  (causal) energy-momentum $n$-point correlators. The transport coefficients $\uplambda_1, \uplambda_2$ and $\uplambda_3$ require 3-point correlators  \cite{Moore:2010bu}. To proceed, one can solve the conservation law $\nabla_\mu T^{ \mu\nu}=0$ in momentum space, by expressing the Fourier transform of Eq. (\ref{npo}) as \cite{Arnold:2011ja}
\bea
\!\!\!\!\!\!\!\!\langle T^{\mu\nu}(\vec{q})\rangle_h&=&\langle T^{\mu\nu}\rangle_{h=0}
-\frac{1}{2(2\pi)^4} \int d^4\vec{q}_1 
\delta^4(\vec{q}-\vec{q}_1) G^{\mu\nu,\sigma\rho}(\vec{q};-\vec{q}_1)h_{\sigma\rho}(\vec{q}_1)\nn\\
&&+ \frac{1}{8(2\pi)^4} \int d^4\vec{q}_1 \int d^4\vec{q}_2 
\delta^4(\vec{q}-\vec{q}_1-\vec{q}_2) G^{\mu\nu,\sigma\rho,\tau\zeta}(\vec{q};-\vec{q}_1,-\vec{q}_2)h_{\sigma\rho}(\vec{q}_1)h_{\tau\zeta}(\vec{q}_2)+\dots \label{npo FT}
\eea 
Regarding a 3-point function, one can assume $
q_1^\mu=(\omega_1,0,0,\mathsf{q}_1)$ and $ q_2^\mu=(\omega_2,0,0,\mathsf{q}_2)$. 
Therefore, the transport coefficients  $\uplambda_1,\uplambda_2$, and $\uplambda_3$ can be read off from 3-point  correlators, in the strong-coupling regime, as \cite{Arnold:2011ja,Bhattacharyya:2007vjd}
\beq
&&\lim_{\substack{\omega_2\to0 \\ \omega_1\to 0}}
\frac{\partial}{\partial{\omega_1}}\frac{\partial}{\partial{\omega_2}} \lim_{\substack{\mathsf{q}_1\to0 \\ \mathsf{q}_2\to 0}}G^{xy,xz,yz}=\upeta\tau_\Uppi-\uplambda_1\nn\\
&&\lim_{\substack{\mathsf{q}_2\to 0\\\omega_1\to 0}}\frac{\partial}{\partial{\mathsf{q}_2}}\frac{\partial}{\partial{\omega_1}}\lim_{\substack{\mathsf{q}_1\to 0\\\omega_2\to 0}}G^{xy,yz,0x}=\frac 12\upeta\tau_{\Pi}-\frac 14\uplambda_2\nn\\
&&\lim_{\substack{\mathsf{q}_2\to0\\\mathsf{q}_1\to0}}
     \frac{\partial}{\partial{\mathsf{q}_1}} \frac{\partial}{\partial{\mathsf{q}_2}}
      \lim_{\substack{\omega_2\to0\\\omega_1\to0}}
            G^{xy,0x,0y}
   =
   -\frac14 \uplambda_3,\label{preview}
\eeq
implying that some of the following transport coefficients carry corrections by quantum gravity, encoded into the parameter $\gamma$:
\begin{subequations}
\beq
\tau_\Uppi&=&\frac{(2-\log 2)}{2\pi T}(1+24\gamma^2)\\
\uplambda_1&=&\frac{N_c^2 T^2}{16}(1+52\gamma^2+16\gamma^4),\\
\uplambda_2 &=& -(1+12\gamma^2+2\gamma^4)\frac{\log 2}{8}N_c^2 T^2,\\ \uplambda_3&=&0,\label{preview2}
\eeq
\end{subequations}
where $T$ denotes the temperature of the system.

To relate transport coefficients to thermal correlators,
the fluid response due to small  perturbations of type $h_{xy}=h_{xy}(t,z)$ must be analyzed.
The effective  energy-momentum  can thus be written as 
\begin{equation}
  (T^\text{eff})^{xy}= -P^\text{eff} h_{xy} -\upeta \partial_t h_{xy} + \upeta\tau_\Uppi\partial^2_t h_{xy}
          -\frac\upkappa 2 \left(\partial^2_t + \partial_z^2\right) h_{xy}\, .
\end{equation}
The linear response theory implies that the retarded correlator  reads
\beq
  G_R^{xy,xy}(\omega, q) &=& P^\text{eff} -i\upeta\omega+\upeta\tau_\Uppi\omega^2
     -\frac\upkappa 2 (\omega^2+q^2).
\label{hydroxyxycorr}
\eeq
In this sense one can write Eq. (\ref{kk}). 
The sound pole can be now computed, once  conformal 
hydrodynamics in stationary equilibrium is considered at an initial time, implying  homogeneous energy density
$\upepsilon^\text{eff} \sim T^4$ and $(\Uppsi^\text{eff})^{\mu\nu}=0$. One then  marginally perturbs  the fluid mechanical system, denoting 
an out-of-equilibrium shift, with effective energy density, $\delta\upepsilon^\text{eff}$, fluid flow velocity, $u^i$, and dissipative spatial part of the energy-momentum tensor, $(\Uppsi^\text{eff})^{ij}$.
For small perturbations, one can neglect the nonlinear terms in
Eq.~(\ref{gr1}) encompassing sound waves, since they are nonlinear functions of the velocity. Linear approximation in the perturbations yields 
\begin{subequations}
\beq
\delta T^{00}&=&\delta \upepsilon^\text{eff},\\\delta T^{0i}&=&\left(\upepsilon^\text{eff}+P^\text{eff}\right) u^i,\\
\delta T^{ij}&=&(c_s^{\rm eff})^2 \delta \upepsilon^\text{eff}\ \delta^{ij}+ \left(\Uppsi^\text{eff}\right)^{ij},
\eeq
\end{subequations}
where the speed of sound, at constant temperature, can be computed by
\begin{equation}
    c_s^\text{eff}=\frac{dP^\text{eff}}{d\upepsilon^\text{eff}} = c_s \ ,
\end{equation}
with $c_s=dP/d\upepsilon$ the classical speed of sound.
For sound waves traveling along the $x$-direction,  energy-momentum  conservation implies
\begin{align}\label{eq:deltaepsilon}
\partial_t (\delta\upepsilon^\text{eff})+(\upepsilon^\text{eff}+P^\text{eff})\partial_x u^x &=0\, ,\\
\label{eq:deltau}
\left(\upepsilon^\text{eff}+P^\text{eff}\right) \partial_t u^x+c_s^2\partial_x \left(\delta\upepsilon^\text{eff}\right)+
\partial_x \left(\Uppsi^\text{eff}\right)^{x x}&=0\, .
\end{align}
Eq.~(\ref{gr1}) can be written as 
\begin{equation}\label{eq:deltaPi}
(\tau_\Uppi \partial_t+1) \left(\Uppsi^\text{eff}\right)^{xx}+4 \upeta \partial_x u^x=0\, .
\end{equation}
Considering a plane wave, Eqs.  (\ref{eq:deltaepsilon}) --  (\ref{eq:deltaPi}) provide the dispersion relation
\begin{equation}\label{eqq}
  - \tau_\Uppi\omega^3 - i \omega^2+q^2 c_s^2 \tau_\Uppi\omega 
  +4q^2 \frac{\upeta}{\upepsilon^\text{eff}+P^\text{eff}}\omega +i c_s^2q^2 =0.
\end{equation}
At the small-$q$ limit, two solutions of Eq. (\ref{eqq}) representing 
sound waves read
\begin{equation}\label{dispsound1}
  \omega_{\pm} = \pm c_s q - i\Upgamma^\text{eff} q^2 \pm \frac{\Upgamma^\text{eff}}{c_s} 
  \left( c_s^2\tau_\Uppi - \frac{\Upgamma^\text{eff}}{2}\right) q^3
+{\cal O}(q^4)\,,
\end{equation}
where the effective diffusion
constant for shear fluctuations is given by 
\begin{equation}
    \Upgamma^\text{eff} = 
    2\frac{\upeta}
    {\upepsilon^\text{eff}+P^\text{eff}}
    =
    \Upgamma
    \ ,
    \label{dc}
\end{equation}
where we used $\upepsilon^\text{eff}+P^\text{eff} = \upepsilon+P$ from 
Eqs.~(\ref{em2}, \ref{peff}) and $\Upgamma$ is the classical diffusion constant.
The additional solution of Eq. \eqref{eqq}, 
\begin{equation}
  \label{eq:sound-omega3}
  \omega=-i\tau_\Uppi^{-1}+\mathcal{O}(q^2)\,,
\end{equation}
 does not equal zero in the limit $q\to0$, indicating a residual macroscopic scale beyond the  hydrodynamical regime.
 
In the functional measure setup,  one can search for the
eigenmodes of the linearized hydrodynamic equations, to understand the locations of the correlators poles. The equation governing the charge diffusion,
\beq\left(\partial_t- D^\text{eff}\nabla^2\right)\uprho = 0,
\eeq for $\uprho$ denoting the electric charge, represents a   pole in the 2-point correlation function at the values $\omega = -iD^\text{eff}q^2$.
To obtain the poles one can, for the sake of simplicity, assume $\vec{q} = q\hat{z}$. Shear eigenmodes can be thus calculated when fluctuations of pairs of components, let us say $T^{0k}$ and $T^{3k}$, are taken into account, where $k =
1, 2$. The constitutive equation reads 
\beq T^{3k} = -\upeta^\text{eff} \del_z u^k = -\frac{\upeta^\text{eff}}{\upepsilon^\text{eff}+P^\text{eff}}\del_z T^{0k},\eeq
whereas the diffusion equation for $T^{0k}$ reads 
\beq 
\left(\del_t+\frac{\upeta^\text{eff}}{\upepsilon^\text{eff}+P^\text{eff}}\del_z\right)T^{0k} =0.\label{diff}\eeq
 Replacing the plane wave $e^{-i\omega t+iqz}$ into (\ref{diff})  yields 
the dispersion law
$\omega^\text{eff} = -i\frac{\upeta}{\upepsilon^\text{eff}+P^\text{eff}}$. The effective diffusion constant 
\beq
D_\upeta^\text{eff} = \frac{\upeta}
{\upepsilon + P}
\eeq
in Eq. (\ref{dc}) mimics the coefficient of the correlator, with  a diffusive pole $ω^\text{eff} = -i\frac{\upeta}{\upepsilon^\text{eff}+P^\text{eff}}q$ \cite{Son:2007vk}. We thus see that $D_\upeta^{\rm eff} = D_\upeta$ does not acquire quantum corrections from the measure.
In the non-relativistic limit, when $P^\text{eff}\ll \upepsilon^\text{eff}$, the effective diffusion constant, $D_\upeta^\text{eff}$, equals the kinematic viscosity.

An important additional feature of the functional measure can be addressed. 
There is a phenomenon of  attenuation in sound waves, due to the effective shear and
bulk viscosities. The effective decay rate $\lambda^\text{eff}$ of sound waves with momentum $\vec{q}$ reads \cite{Jeon:1995zm} 
\beq
\lambda^\text{eff} = 
\left(\frac43\upeta + \zeta^\text{eff}\right)
\frac{q^2}{\upepsilon+P}=\lambda+\frac{4i\gamma q^2}{\upepsilon+P},\eeq
expressing a quantum correction to the decay rate  of sound waves, $\lambda$, due to the functional measure.

\begin{figure}[H]\begin{center}
\includegraphics[scale=0.75]{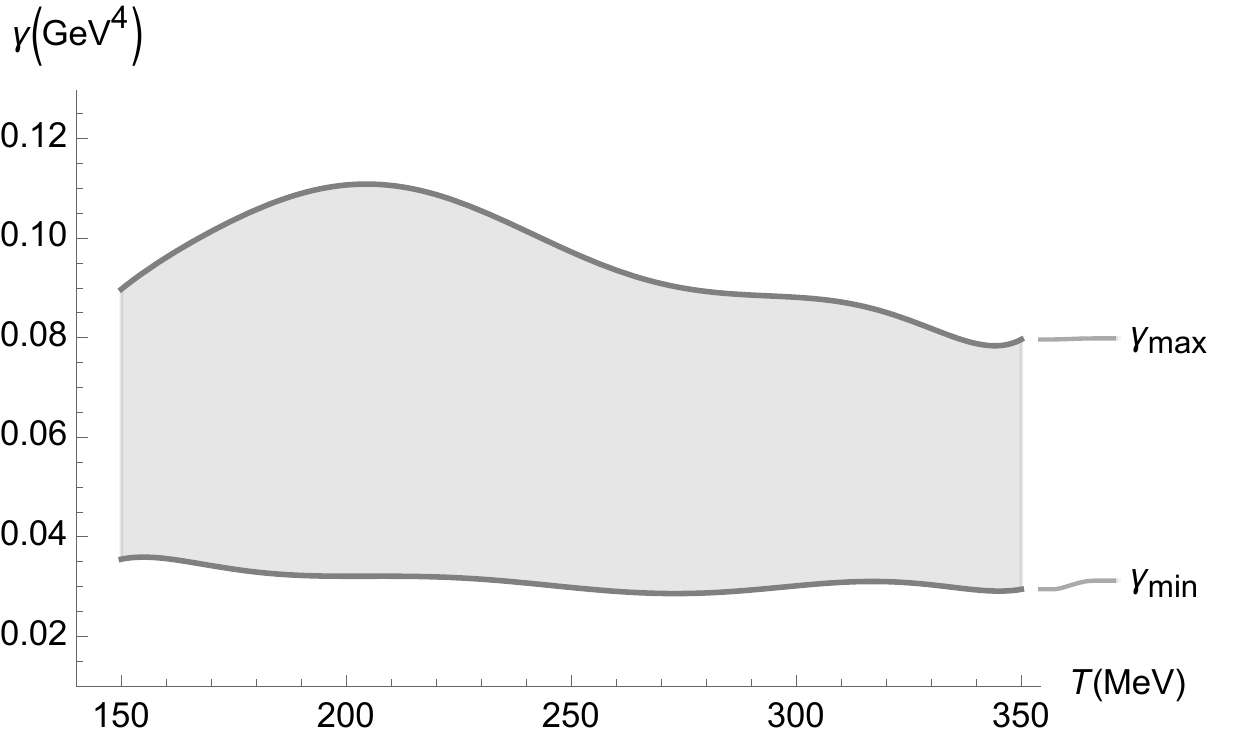}
\caption{The bound $\gamma_{\text{min}} < \gamma < \gamma_{\text{max}}$ (GeV$^4$) as a function of the QGP temperature (MeV), using the results by the  JETSCAPE Bayesian model   \cite{JETSCAPE:2020shq}.}
\label{gmm}\end{center}
\end{figure}

\begin{figure}[H]\begin{center}
\includegraphics[scale=0.75]{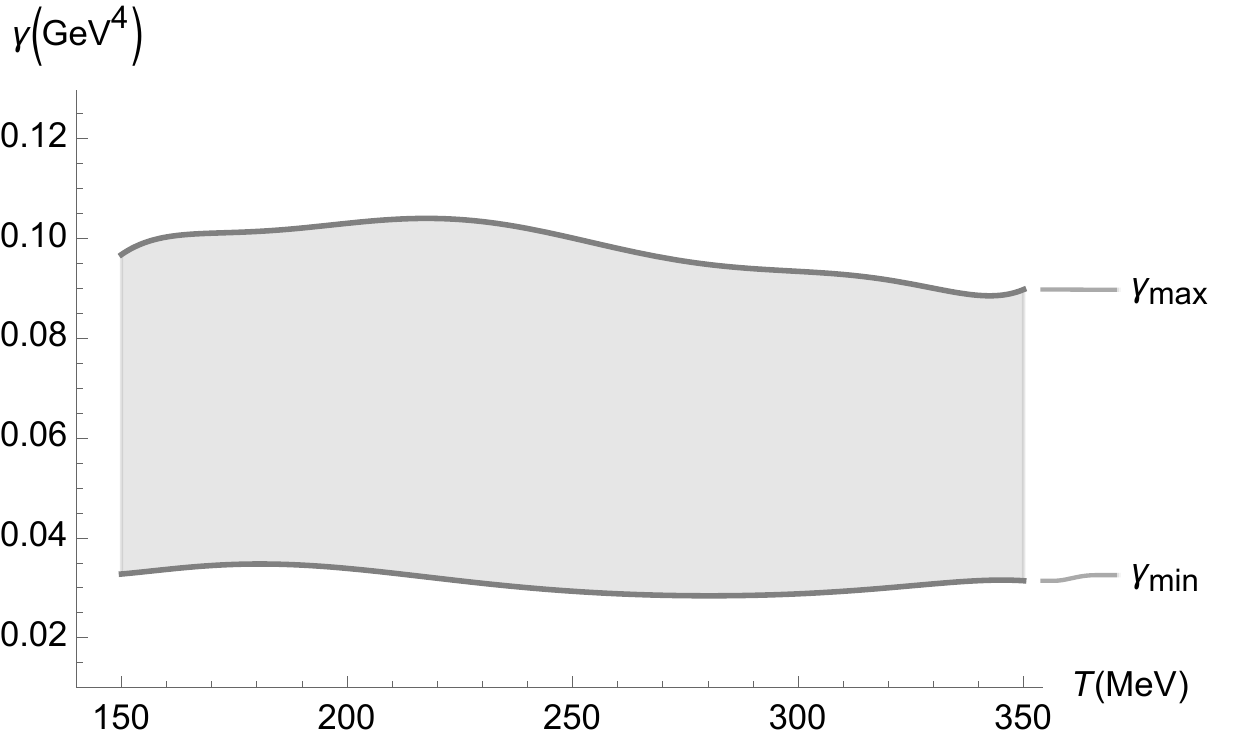}
\caption{The bound $\gamma_{\text{min}} < \gamma < \gamma_{\text{max}}$ (GeV$^4$) as a function of the QGP temperature (MeV), using the measured experimental  values of the bulk viscosity and analysis  by the Duke group \cite{Bernhard:2019bmu}.}
\label{gmm1}\end{center}
\end{figure}

\begin{figure}[H]\begin{center}
\includegraphics[scale=0.75]{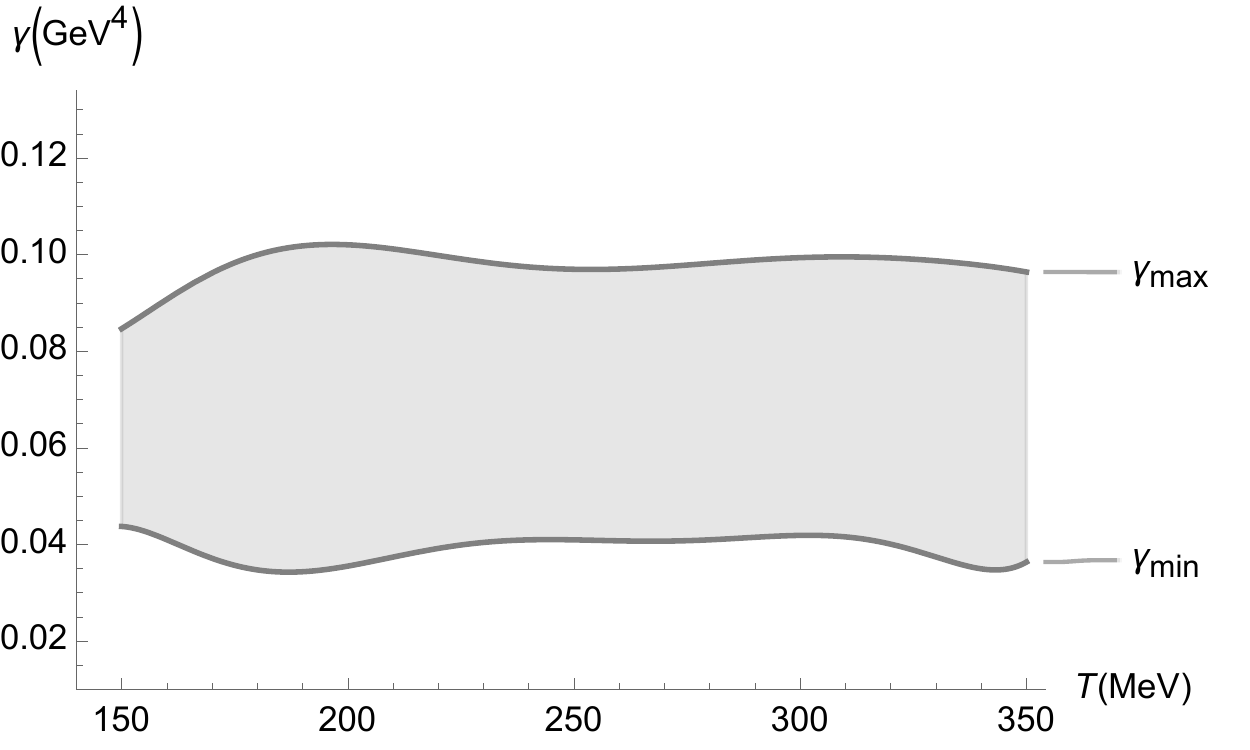}
\caption{The bound $\gamma_{\text{min}} < \gamma < \gamma_{\text{max}}$ (GeV$^4$) as a function of the QGP temperature (MeV), using the measured experimental values of the bulk viscosity and analysis by the Jyv\"askyl\"a-Helsinki-Munich group \cite{Parkkila:2021yha}.}
\label{gmm3}\end{center}
\end{figure}

We can now search for a reliable bound on the parameter $\gamma$, which drives the quantum gravity  corrections due to the functional measure. To accomplish this task, we can compare existing experimental data on transport coefficients of the QGP and compare with the ones whose quantum corrections have been  predicted in Sects. \ref{s3} and \ref{s4}. 
Transport coefficients of QGP have been obtained in heavy-ion collision experiments for temperatures in the range $150\lesssim T\lesssim 350$ MeV.  Therefore, Eq. (\ref{bul}) can be taken into account, denoting \beq
\left|\zeta^\text{eff}\right| = \zeta\sqrt{1-\frac{16\gamma^2}{\zeta}}.
\eeq
We can split the analyzes of the latest robust experimental estimates for the QGP bulk viscosity into three parts. The first one regards the JETSCAPE Bayesian model using  \cite{JETSCAPE:2020shq}, complying with the theoretical results in Refs. \cite{qgp,Critelli:2017oub,Soloveva:2019xph}. Therefore a bound $\gamma_{\text{min}} < \gamma < \gamma_{\text{max}}$ for the parameter $\gamma$, encoding quantum corrections due to the functional measure, is depicted in Fig. \ref{gmm} as a function of the temperature.
\noindent The second similar result can be deployed by considering the experimental data for the bulk viscosity of the QGP, obtained by the Duke group \cite{Bernhard:2019bmu}. This time, the temperature-dependent bound on $\gamma$ is plotted in Fig. \ref{gmm1}.
\noindent Finally, the up-to-date experimental results of the bulk viscosity, by the  Jyv\"askyl\"a-Helsinki-Munich group \cite{Parkkila:2021yha}, can be employed. The associated bound for the parameter $\gamma$, as a function of the temperature, is shown in Fig. \ref{gmm3}. 
\noindent Irrespectively the similar results concerning the parameter $\gamma = \gamma(\Lambda)$ (which runs according to Eq.~\eqref{RGE}) in Figs. \ref{gmm} -- \ref{gmm3}, an energy scale $\Lambda \sim$ 3 TeV can be assumed, since the experimental results involve Pb–Pb collision
data at $\sqrt{s_{NN}}$ = 2.76 and 5.02 TeV.

\section{Concluding remarks}
\label{s5}

The consequences of a  functional measure on the boundary CFT were here analyzed and discussed, which has led to important corrections to transport coefficients in AdS/CFT. The modifications due to a functional measure yield an  effective  energy-momentum tensor, whose temporal component, the effective energy density, and the effective pressure (from the spatial components) both acquired a non-trivial contribution  from
the functional measure. This correction was shown to be imaginary, which suggests an unstable fluid and agrees with the difficulty of creating a long-lasting strongly-coupled fluid, say QGP, in the laboratory.
Quantum corrections from the functional measure were broadly investigated for both perfect and viscous fluid flows, including second-order hydrodynamics.
In particular, the effective shear and bulk viscosities have been computed. While the latter receives a correction, the former does not. This strengthens the result from \cite{Kuntz:2019omq}, thus shear viscosity does not receive any quantum corrections at one-loop. Similarly, the entropy density (hence the shear viscosity-to-entropy density ratio), the speed of sound, and the diffusion constants receive no corrections from the functional measure. The reason boils down to their dependence on the sum $\upepsilon^{\rm eff} + P^{\rm eff}$, whose corrections from the effective energy density and the effective pressure cancel out. We stress, however, that this cancellation only takes place for a Minkowski background. Curved background metrics can generally provide non-trivial corrections. The experimental analysis of fluids over curved substrates, for example, could shed light on the quantum effects of gravity. In particular, such experiments could be used to bound the values of the parameter $\gamma$. 
The decay rate in sound wave attenuation, on the other hand, depends on the bulk viscosity and thus gets modified by the functional measure correction.
Besides, the relaxation time and coefficients of conformal traceless tensor fields in second-order hydrodynamics,  were shown to carry significant quantum corrections due to the functional measure.


As large-$N_c$ Yang--Mills theories can be implemented to describe QCD, the results here obtained can be straightforwardly applied to other aspects of the QGP \cite{qgp}. Experiments in the RHIC have constantly shown that the QGP can be described by hydrodynamics, as a fluid with small viscosity, making it a  strongly-coupled system. The quantum gravity corrections due to a functional measure may be tested, in principle, in experiments involving the QGP at the hydrodynamical regime. Regarding it, an important procedure has been implemented and shown in Figs. \ref{gmm} -- \ref{gmm3}, showing the temperature-dependent  bounds for the parameter $\gamma$, carrying quantum gravity effects on the bulk viscosity, in Minkowski background. The analysis leading to our results was based on the last experimental results about the QGP  \cite{JETSCAPE:2020shq,Bernhard:2019bmu,Parkkila:2021yha}.

Another interesting possibility, as shown in Ref. \cite{Denicol:2020eij}, concerns the fact that  hydrodynamics can be seen as a universal attractor, with specific hydrodynamical signatures remaining, even in extreme behaviors of the QGP, including chiral transport
and anomalies  in the context of a functional measure. 
Considerations of this kind can (and should) be generalized with the inclusion of a functional measure. Its details shall be postponed for some future work.

\subsubsection*{Acknowledgments} 
\noindent RdR~is grateful to The S\~ao Paulo Research Foundation -- FAPESP (Grants No. 2021/01089-1 and No. 2022/01734-7) and
the National Council for Scientific and Technological Development -- 
CNPq (Grant No. 303390/2019-0), for partial financial support. IK also thanks the National Council for Scientific and Technological Development -- 
CNPq (Grant No. 303283/2022-0) for financial support.


\begin{thebibliography}{10}

\bibitem{malda}
O.~Aharony, S.~S.~Gubser, J.~M.~Maldacena, H.~Ooguri and Y.~Oz,
Phys. Rept. \textbf{323} (2000) 183 
[arXiv:hep-th/9905111 [hep-th]].


\bibitem{Nastase:2015wjb}
H.~Nastase,
``\emph{Introduction to the AdS/CFT Correspondence},'' Cambridge University Press, Cambridge, 2015. 

\bibitem{Iqbal:2008by}
N.~Iqbal and H.~Liu,
Phys. Rev. D \textbf{79} (2009) 025023
[arXiv:0809.3808 [hep-th]].

\bibitem{Son:2007vk}
D.~T.~Son and A.~O.~Starinets,
Ann. Rev. Nucl. Part. Sci. \textbf{57} (2007) 95 
[arXiv:0704.0240 [hep-th]].

\bibitem{Bhattacharyya:2007vjd}
S.~Bhattacharyya, V.~E.~Hubeny, S.~Minwalla and M.~Rangamani,
JHEP \textbf{02} (2008) 045
[arXiv:0712.2456 [hep-th]].






\bibitem{Haack:2008cp}
M.~Haack and A.~Yarom,
JHEP \textbf{10} (2008) 063
[arXiv:0806.4602 [hep-th]].


\bibitem{Policastro:2002se}
G.~Policastro, D.~T.~Son and A.~O.~Starinets,
JHEP \textbf{09} (2002) 043
[arXiv:hep-th/0205052 [hep-th]].


\bibitem{Bernardo:2018cow}
H.~Bernardo and H.~Nastase,
JHEP \textbf{12} (2019) 025
[arXiv:1812.07586 [hep-th]].

\bibitem{Policastro:2001yc}
G.~Policastro, D.~T.~Son and A.~O.~Starinets,
Phys. Rev. Lett. \textbf{87} (2001) 081601 
[arXiv:hep-th/0104066 [hep-th]].


\bibitem{Ferreira-Martins:2021cga}
A.~J.~Ferreira-Martins and R.~da Rocha,
Nucl. Phys. B \textbf{973} (2021) 115603
[arXiv:2104.02833 [hep-th]].

\bibitem{Ferreira-Martins:2019svk}
A.~J.~Ferreira\textendash{}Martins, P.~Meert and R.~da Rocha,
Nucl. Phys. B \textbf{957} (2020) 115087
[arXiv:1912.04837 [hep-th]].


\bibitem{Bemfica:2020xym}
F.~S.~Bemfica, M.~M.~Disconzi, V.~Hoang, J.~Noronha and M.~Radosz,
Phys. Rev. Lett. \textbf{126} (2021) 222301
[arXiv:2005.11632 [hep-th]].

\bibitem{Bemfica:2017wps}
F.~S.~Bemfica, M.~M.~Disconzi and J.~Noronha,
Phys. Rev. D \textbf{98} (2018)  104064
[arXiv:1708.06255 [gr-qc]].


\bibitem{Rocha:2022ind}
G.~S.~Rocha, G.~S.~Denicol, J.~Noronha,
Phys. Rev. D \textbf{106} (2022)  036010
[arXiv:2205.00078 [nucl-th]].

\bibitem{Kovtun:2019hdm}
P.~Kovtun,
JHEP \textbf{10} (2019) 034
[arXiv:1907.08191 [hep-th]].

\bibitem{Hoult:2020eho}
R.~E.~Hoult and P.~Kovtun,
JHEP \textbf{06} (2020) 067
[arXiv:2004.04102 [hep-th]].

\bibitem{Meert:2018qzk}
P.~Meert and R.~da Rocha,
Eur. Phys. J. C \textbf{78} (2018)  1012
[arXiv:1809.01104 [hep-th]].


\bibitem{Bonora:2014dfa}
L.~Bonora, K.~P.~S.~de Brito and R.~da Rocha,
JHEP \textbf{02} (2015) 069
[arXiv:1411.1590 [hep-th]].

\bibitem{DerradideSouza:2015kpt}
R.~Derradi de Souza, T.~Koide,  T.~Kodama,
Prog. Part. Nucl. Phys. \textbf{86} (2016) 35 
[arXiv:1506.03863 [nucl-th]].

\bibitem{Karapetyan:2018yhm}
G.~Karapetyan,
Phys. Lett. B \textbf{786} (2018) 418 
[arXiv:1807.04540 [nucl-th]].

\bibitem{Kovtun:2004de}
P.~Kovtun, D.~T.~Son and A.~O.~Starinets,
Phys. Rev. Lett. \textbf{94} (2005) 111601
[arXiv:hep-th/0405231 [hep-th]].

\bibitem{Kuntz:2019omq}
I.~Kuntz and R.~da Rocha,
Nucl. Phys. B \textbf{961} (2020) 115265
[arXiv:1909.10121 [hep-th]].

\bibitem{Montenegro:2020paq}
D.~Montenegro and G.~Torrieri,
Phys. Rev. D \textbf{102} (2020) 036007
[arXiv:2004.10195 [hep-th]].

\bibitem{Vilkovisky:1984st}
G.~A.~Vilkovisky,
Nucl. Phys. B \textbf{234} (1984) 125 

\bibitem{Mottola:1995sj}
E.~Mottola,
J. Math. Phys. \textbf{36} (1995) 2470 
[arXiv:hep-th/9502109 [hep-th]].

\bibitem{Fujikawa:1983im}
K.~Fujikawa,
Nucl. Phys. B \textbf{226} (1983) 437

\bibitem{Fujikawa:1979ay}
K.~Fujikawa,
Phys. Rev. Lett. \textbf{42} (1979) 1195 

\bibitem{Toms:1986sh}
D.~J.~Toms,
Phys. Rev. D \textbf{35} (1987) 3796

\bibitem{DeWitt:2003pm}
B.~S.~DeWitt,
``\emph{The global approach to quantum field theory}'', Oxford University Press, Oxford, 2003.

\bibitem{Casadio:2022ozp}
R.~Casadio, A.~Kamenshchik and I.~Kuntz,
[arXiv:2210.04368 [hep-th]].

\bibitem{Meetz:1969as}
K.~Meetz,
J. Math. Phys. \textbf{10} (1969) 589.

\bibitem{Slavnov:1971mz}
A.~A.~Slavnov and L.~D.~Faddeev,
Teor. Mat. Fiz. \textbf{8} (1971)  297.



\bibitem{Fradkin:1973wke}
E.~S.~Fradkin and G.~A.~Vilkovisky,
Phys. Rev. D \textbf{8} (1973) 4241. 

\bibitem{Fradkin:1976xa}
E.~S.~Fradkin and G.~A.~Vilkovisky,
Lett. Nuovo Cim. \textbf{19} (1977) 47. 

\bibitem{Casadio:2021rwj}
R.~Casadio, A.~Kamenshchik and I.~Kuntz,
Nucl. Phys. B \textbf{971} (2021), 115496
[arXiv:2102.10688 [hep-th]].

\bibitem{Kuntz:2022tat}
I.~Kuntz, R.~Casadio and A.~Kamenshchik,
Mod. Phys. Lett. A \textbf{37} (2022) no.10, 2230007
[arXiv:2203.11259 [hep-th]].

\bibitem{Casadio:2020zmn}
R.~Casadio, A.~Kamenshchik and I.~Kuntz,
Int. J. Mod. Phys. D \textbf{31} (2022) no.01, 2150130
[arXiv:2008.09387 [gr-qc]].

\bibitem{Witten:1998qj}
E.~Witten,
Adv. Theor. Math. Phys. \textbf{2} (1998) 253 
[arXiv:hep-th/9802150 [hep-th]].


\bibitem{Gubser:1998bc}
S.~S.~Gubser, I.~R.~Klebanov, A.~M.~Polyakov,
Phys. Lett. B \textbf{428} (1998) 105 
[arXiv:hep-th/9802109 [hep-th]].

\bibitem{Natsuume:2014sfa}
M.~Natsuume,
Lect. Notes Phys. \textbf{903} (2015) 1
[arXiv:1409.3575 [hep-th]].


\bibitem{Casadio:2016zhu}
R.~Casadio, R.~T.~Cavalcanti and R.~da Rocha,
Eur. Phys. J. C \textbf{76} (2016) 556
[arXiv:1601.03222 [hep-th]].

\bibitem{Torrieri:2007fb}
G.~Torrieri, B.~Tomasik and I.~Mishustin,
Phys. Rev. C \textbf{77} (2008) 034903 
[arXiv:0707.4405 [nucl-th]].

\bibitem{Finazzo:2014cna}
S.~I.~Finazzo, R.~Rougemont, H.~Marrochio and J.~Noronha,
JHEP \textbf{02} (2015)  051
[arXiv:1412.2968 [hep-ph]].


\bibitem{Gubser:2008sz}
S.~S.~Gubser, S.~S.~Pufu and F.~D.~Rocha,
JHEP \textbf{08} (2008)  085
[arXiv:0806.0407 [hep-th]].


\bibitem{Buchel:2007mf}
A.~Buchel,
Phys. Lett. B \textbf{663} (2008) 286 
[arXiv:0708.3459 [hep-th]].


\bibitem{Buchel:2016cbj}
A.~Buchel, M.~P.~Heller and J.~Noronha,
Phys. Rev. D \textbf{94} (2016) 106011
[arXiv:1603.05344 [hep-th]].





\bibitem{Arnold:2011ja}
P.~Arnold, D.~Vaman, C.~Wu and W.~Xiao,
JHEP \textbf{10} (2011) 033
[arXiv:1105.4645 [hep-th]].

\bibitem{Kovtun:2011np}
P.~Kovtun, G.~D.~Moore and P.~Romatschke,
Phys. Rev. D \textbf{84} (2011) 025006
[arXiv:1104.1586 [hep-ph]].
\bibitem{Montenegro:2016gjq}
D.~Montenegro and G.~Torrieri,
Phys. Rev. D \textbf{94} (2016)  065042
[arXiv:1604.05291 [hep-th]].

\bibitem{Schenke:2012wb}
B.~Schenke, P.~Tribedy, R.~Venugopalan,
Phys. Rev. Lett. \textbf{108} (2012)  252301
[arXiv:1202.6646 [nucl-th]].


\bibitem{GoncalvesdaSilva:2017bvk}
A.~Goncalves da Silva and R.~da Rocha,
Phys. Lett. B \textbf{774} (2017) 98 
[arXiv:1706.01482 [hep-ph]].


\bibitem{Oliveira:2020yac}
O.~Oliveira, T.~Frederico and W.~de Paula,
Eur. Phys. J. C \textbf{80} (2020)  484
[arXiv:2006.04982 [hep-ph]].


\bibitem{Braga:2022yfe}
N.~R.~F.~Braga, L.~F.~Faulhaber and O.~C.~Junqueira,
Phys. Rev. D \textbf{105} (2022)  106003
[arXiv:2201.05581 [hep-th]]. 

\bibitem{Romatschke:2009kr}
P.~Romatschke,
Class. Quant. Grav. \textbf{27} (2010)  025006
[arXiv:0906.4787 [hep-th]]. 



\bibitem{Baier:2007ix}
R.~Baier, P.~Romatschke, D.~T.~Son, A.~O.~Starinets and M.~A.~Stephanov,
JHEP \textbf{04} (2008) 100
[arXiv:0712.2451 [hep-th]].


\bibitem{Natsuume:2007ty}
M.~Natsuume and T.~Okamura,
Phys. Rev. D \textbf{77} (2008), 066014
[erratum: Phys. Rev. D \textbf{78} (2008) 089902]
[arXiv:0712.2916 [hep-th]].

\bibitem{Rocha:2022fqz}
G.~S.~Rocha, M.~N.~Ferreira, G.~S.~Denicol and J.~Noronha,
Phys. Rev. D \textbf{106} (2022)   036022
[arXiv:2203.15571 [nucl-th]].

\bibitem{Rocha:2021zcw}
G.~S.~Rocha, G.~S.~Denicol, J.~Noronha,
Phys. Rev. Lett. \textbf{127} (2021)  042301
[arXiv:2103.07489 [nucl-th]].
\bibitem{Moore:2010bu}
G.~D.~Moore and K.~A.~Sohrabi,
Phys. Rev. Lett. \textbf{106} (2011) 122302
[arXiv:1007.5333 [hep-ph]].

\bibitem{Jeon:1995zm}
S.~Jeon and L.~G.~Yaffe,
Phys. Rev. D \textbf{53} (1996) 5799 
[arXiv:hep-ph/9512263 [hep-ph]].



\bibitem{JETSCAPE:2020shq}
D.~Everett \textit{et al.} [JETSCAPE],
Phys. Rev. Lett. \textbf{126} (2021)  242301
[arXiv:2010.03928 [hep-ph]].




\bibitem{qgp}
J.~Grefa, M.~Hippert, J.~Noronha, J.~Noronha--Hostler, I.~Portillo, C.~Ratti and R.~Rougemont,
Phys. Rev. D \textbf{106} (2022)  034024
[arXiv:2203.00139 [nucl-th]].

\bibitem{Critelli:2017oub}
R.~Critelli, J.~Noronha, J.~Noronha-Hostler, I.~Portillo, C.~Ratti and R.~Rougemont,
Phys. Rev. D \textbf{96} (2017)  096026
[arXiv:1706.00455 [nucl-th]].

\bibitem{Soloveva:2019xph}
O.~Soloveva, P.~Moreau, E.~Bratkovskaya,
Phys. Rev. C \textbf{101} (2020) 045203
[arXiv:1911.08547 [nucl-th]].

\bibitem{Bernhard:2019bmu}
J.~E.~Bernhard, J.~S.~Moreland and S.~A.~Bass,
Nature Phys. \textbf{15} (2019)  1113.


\bibitem{Parkkila:2021yha}
J.~E.~Parkkila, A.~Onnerstad, S.~F.~Taghavi, C.~Mordasini, A.~Bilandzic, M.~Virta and D.~J.~Kim,
Phys. Lett. B \textbf{835} (2022) 137485
[arXiv:2111.08145 [hep-ph]].

\bibitem{Denicol:2020eij}
G.~S.~Denicol and J.~Noronha,
Nucl. Phys. A \textbf{1005} (2021)  121748
[arXiv:2003.00181 [nucl-th]].









\end{thebibliography}
\end{document}